# Discommensurational and Inhomogeneous States Induced by a Strong Magnetic Field in Low-Dimensional Antiferromagnets

B.A. Ivanov [1], C.E. Zaspel [2], A.Yu. Merkulov [1]

[1] Institute of Magnetism, NASU, Vernadskii Ave. 36-B, 03142, Kiev, Ukraine
[2] Department of Physics, Montana State University-Bozeman, Bozeman, MT 59717, USA

Abstract

Anisotropic Antiferromagnetic systems of dimensionality greater than one in an external field are shown to exhibit a complicated array of ground states depending on the spin structure of the surface. The simplest structure that exhibits these effects is the spin ladder with the surface being the ladder end, which can be either compensated or non-compensated spins. The structure with the compensated end has a surface spin flop phase, the non-compensated end has a discommensurational phase, and the transition to these phases can be either first or second order with a tricritical point.

Corresponding Author: C. E. Zaspel,
email: zaspel@physics.montana.edu
Tel: 406-994-3614
Fax: 406-994 4452

Spin-flop transitions in antiferromagnetic (AFM) systems induced by magnetic fields have been studied for more than fifty years [1-3], and this area still generates much interest. Using the classical semi-infinite spin chain model with single-ion anisotropy, Mills mentioned [3] that spin flop states could be localized at the surface of an AFM system at a critical field that is lower than the bulk critical field. This state was described by. This surface spin flop (SSF) state recently was observed [4,5] in Fe/Cr multilayer systems, which stimulated renewed interest in the nature of localized surface states [4-8]. In all of these references the one-dimensional model has been used since it is obviously adequate [3] for multilayer systems, as well as for AFM systems with a simple surface consisting of spins from one sublattice only.

In this letter the nature of the surface states and the transitions to these states are investigated for systems of dimensionality greater than one. Even for the simple spin ladder, which is intermediate between one and two-dimensional systems there are unexpected and interesting effects absent from the simple spin chain model considered previously [3, 6-8]. These effects originate in the more complicated surfaces that are possible in ladder structures and in two-dimensional arrays. In general there can be two types of surfaces (compensated, with an equal number of spins from different sublattices on the surface and zero surface magnetization in the AFM state; and non-compensated, with nonzero surface magnetization). For the last case, with the non-compensated end spin antiparallel to the external field, the discommensuration state with a $180^0$ domain wall common to that considered for spin chains [6-8] appears. In contrast, for compensated surfaces the SSF phase, as proposed by Mills [3], is found. In this state the surface spins rotate to about $90^0$ from the external field with a net magnetic moment at the surface, and there is uniform decay to the AFM phase moving into the bulk. For both cases a critical field is lower than the bulk critical field.

In past investigations of chain models only first order transitions were reported in the literature [6-8]. However, for these more complicated structures it is shown here that the transition to the SSF phase or discommensurational states can be either first or second order depending on the type of anisotropy (exchange or single-ion anisotropy) as well as on the character (compensated or noncompensated) of the surface spins. For the second order transition the amplitude of the nonuniform spin distribution goes to zero at the transition point. In the vicinity of this point the state is neither SSF nor discommensurational, rather it is a slightly broken Neel state with the deviation from the Neel state decaying into the bulk with oscillations. Moreover, with both types of anisotropy, a tricritical point can be present.

Both numerical and analytical methods are used to investigate the nature of the ground state with different surfaces and anisotropies. We begin with the discrete Hamiltonian of the uniaxial spin ladder with classical spins $\mathbf{S}_\mathbf{i}$ at sites $\mathbf{i}$ of a dimer lattice and an AFM interaction between nearest neighbours,

$$W=\sum_{\mathbf{i},\delta}\left(J\mathbf{S}_\mathbf{i}\mathbf{S}_{\mathbf{i}+\delta}+\kappa S_{z,\mathbf{i}}S_{z,\mathbf{i}+\delta}\right)-k\sum_{\mathbf{i}}S_{z,\mathbf{i}}^2-H\sum_{\mathbf{i}}S_{z,\mathbf{i}}\ . \qquad (1)$$

Here, the first term describes the interaction of neighbouring spins connected by the vector $\delta$, $J$ is the exchange integral, $\kappa$ is the measure of the exchange anisotropy, $k$ is the single-ion anisotropy constant, and $H$ is the external field along the easy $z$-axis in energy units $g\mu_B$, where $g$ is gyromagnetic ratio and $\mu_B$ is the Bohr magneton. For description of the ladder system it is natural to take a single dimer as the magnetic unit cell, and use the net magnetization, $\mathbf{m}_n$ and the antiferromagnetic vector, $\mathbf{l}_n$ for $n$-th dimer,

$$\mathbf{m}_n = (\mathbf{S}_1 + \mathbf{S}_2)/2S\ ,\ \mathbf{l}_n = (\mathbf{S}_1 - \mathbf{S}_2)/2S\ . \qquad (2)$$

It is sufficient to consider the spins confined to one plane and to express $\mathbf{m}_n$ and $\mathbf{l}_n$ through the angular variable $\theta_n$ and the length of the magnetization, $m_n$

$$l_z = l\cos\theta,\ \ l_x = l\sin\theta\ ,\ m_z = m\sin\theta\ ,$$

$$m_x = -m\cos\theta\ , \qquad l=\sqrt{1-m^2}\ . \qquad (3)$$

Elimination of $m$ for the infinite antiferromagnet without a surface gives the energy $W(\theta)$ as a function of $\theta$. In the lowest approximation in the small parameters $k/J$ and $\kappa/J$, one can find the effective magnetic anisotropy per dimer, $W(\theta) = K\sin^2\theta$, $K = S^2(2k + Z\kappa)$, which is easy-axis for $K > 0$, and $Z$ is the coordination number. (In the following we put $S = 1$.) It follows that the collinear Neel state, in which $\theta = 0$ or $\pi$ is stable for $H < H_1$, where

$$H_1 = \sqrt{K(2ZJ + K)}, \qquad (4)$$

and a spin-flop phase, with $\theta = \pm\pi/2$, of lower energy than the Neel state is stable for $H > H_{SF} = \sqrt{K(2ZJ + Z\kappa - 2k)}$. If $k > 0$, then $H_{SF} < H_1$ and the spin flop transition is of first order. As will be shown below, the second order transition is the more common case for the surface spin flop transition.

The description of surface phase transitions begins with the numerical minimization of the discrete Hamiltonian for the two most interesting cases: the regular ladder (RL) structure that is a semi-infinite spin ladder having a regular dimer at the end, and the ladder with the single noncompensated (NC) spin on the end. (The configurations of atoms for these cases are present on the inserts in the figures below.) The energy minimization has been performed through a Seidel-like algorithm, for spin ladders as long as 100 dimers, which is much larger that the size of local state. The spins on one end of the ladder are free, and with spins on other end are fixed in the Neel state, corresponding to the bulk non-perturbed state. The distributions of $m_z$ and $l_z$ as a function of the distance from the end of the ladder are qualitatively different for the RL and NC structures as can be seen in Fig.1.

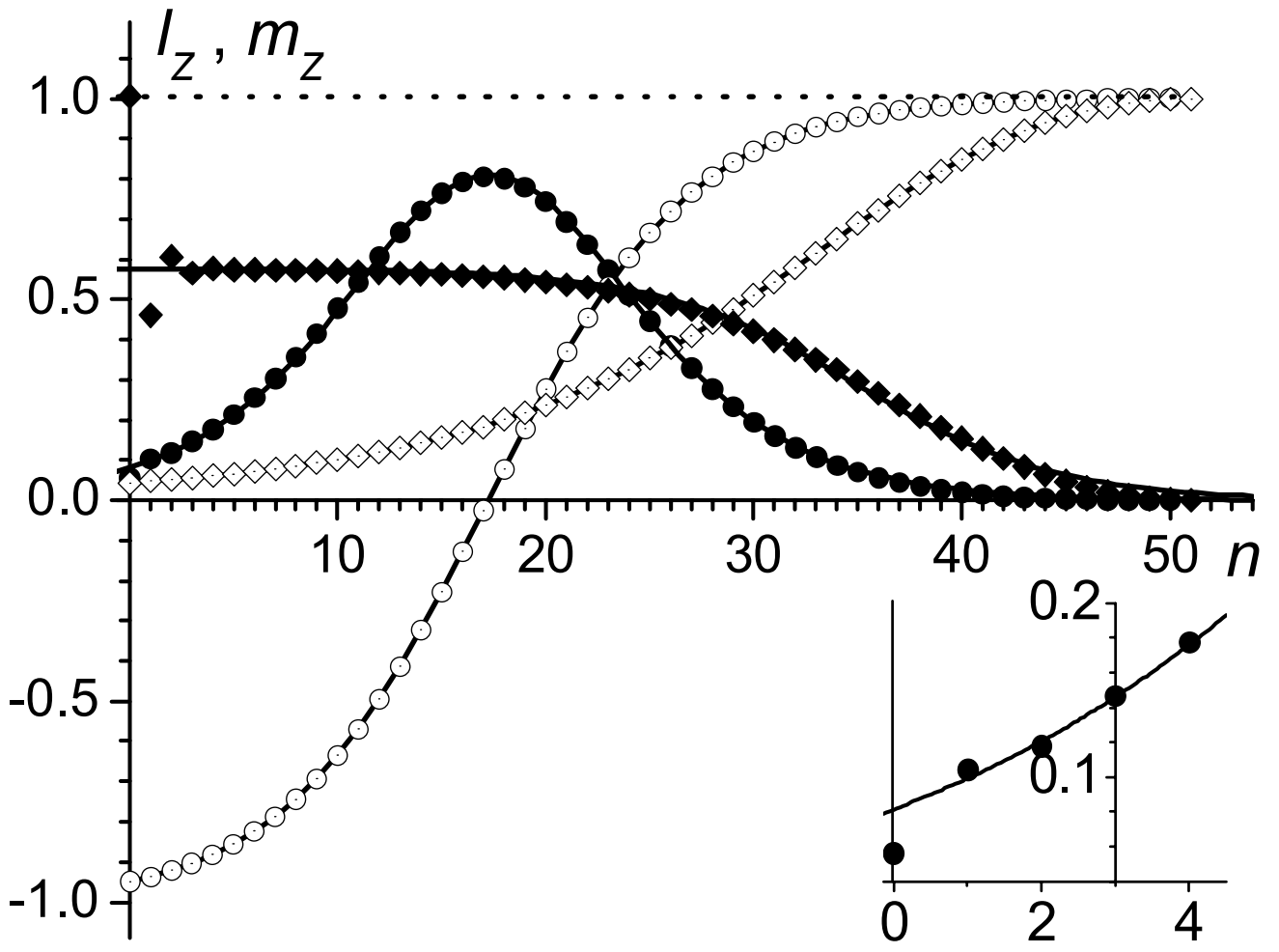

Fig. 1. The spin distribution for different ladder systems with single-ion anisotropy for magnetic fields (in units of $J$) far below the critical field; circles - regular ladder, $H = 0.848$, $k$=0.06, diamonds – NC ladder, $H$=0.24, $k$=0.01. Open symbols represent $l_z$ , full symbols represent $m_z$ , multiplied by 4 for NC ladder and by 20 for the RL. For the last case the end details are shown in the insert.

For the RL structure, which is a model for the spin flop transition in AFM systems with a compensated infinite surface, the $l_z$ data show a surface spin-flop state described by Mills [3] where $\boldsymbol{l}$ rotates approximately $90^0$ with a magnetization that decays to zero into the ladder. For the NC ladder, the discommensuration state [6-8] containing a 180-degree domain wall appears. The $l_{z,\ n}$ data for both cases are well-described by AFM domain wall in the continuum approximation [2]: a 90-degree domain wall with $\tan\theta_n = \exp[-(n-n_0)/\Delta_{\pi/2}]$, or a 180-degree domain wall with $\tan(\theta_n/2) = \exp[-(n-n_0)/\Delta_\pi]$. Numerical data illustrate the non-regular behaviour of $m_z$ near the end of ladder.

The numerical analysis also shows the presence of two types of behaviour for the dependence of the $z$ component of total spin $S_z$ on the magnetic field. For both systems with single-ion anisotropy, the nonuniform phases have a finite value of $S_z$ at the transition point, where the energies of collinear and nonuniform phases coincide. However, for the case of exchange anisotropy the value of $S_z$ goes to zero at the transition point as shown in Fig. 2. This can be interpreted as the presence of first and second order transitions, respectively.

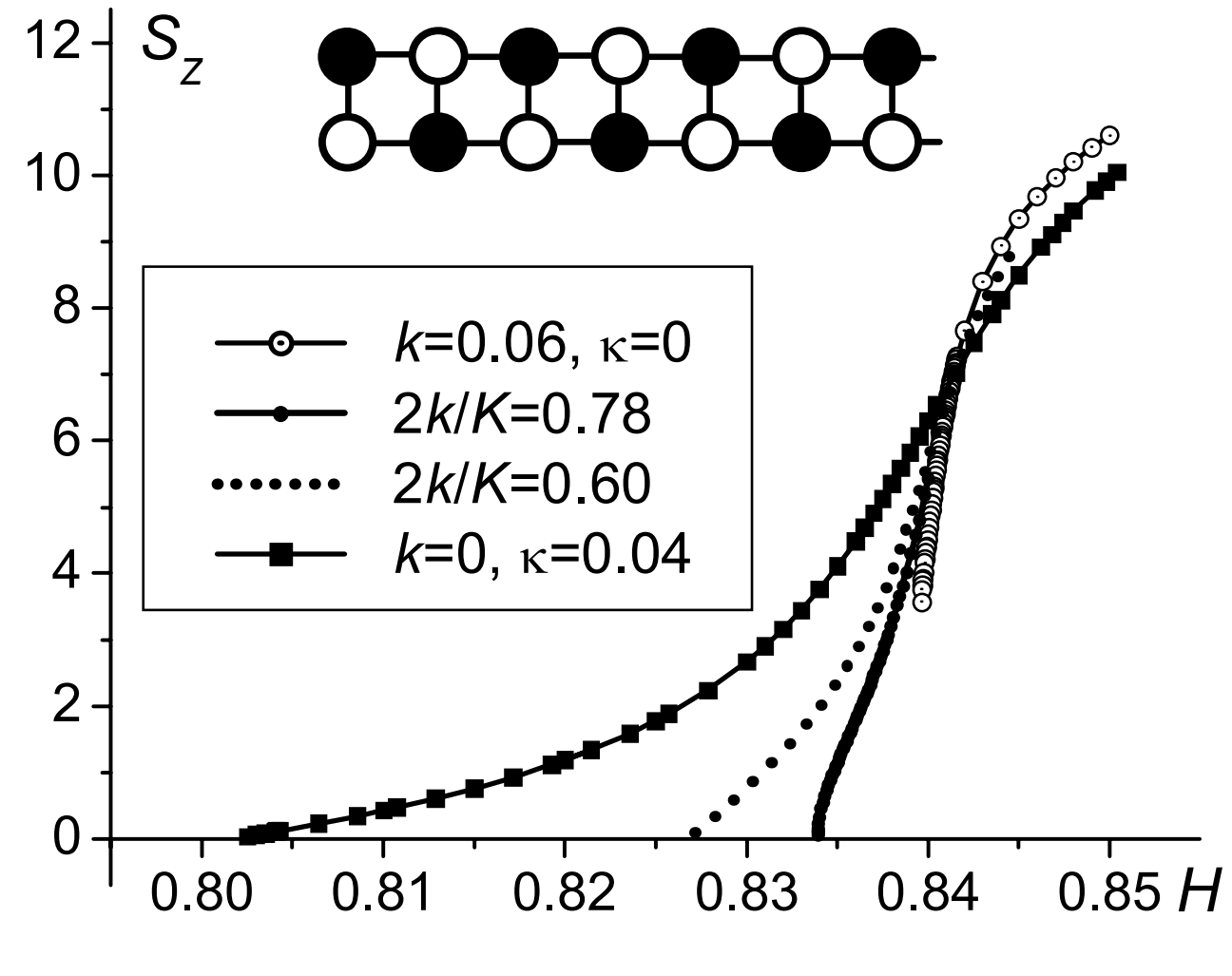

Fig.2. The $S_z(H)$ dependence for regular ladder with single-ion anisotropy, exchange anisotropy and two types of combined anisotropy. For the spin ladder structure shown here and in Fig. 3, open and full circles presents up and down spins in the Neel state, respectively.

To explain these features, consider the stability problem of the collinear phase, having the values of $\theta_n = 0$, and $m_n = 0$. The stability of this state can be investigated using the quadratic approximate Hamiltonian with small variables $\theta_n$ and $m_n$ written as $W = W^{(r)} + \Delta W$, where $W^{(r)}$ is for the regular semi-infinite ladder,

$$W^{(r)} = \sum_{n=0}^{\infty} \left[ (3J + 2k + 3\kappa) m_n^2 + 2J m_n m_{n+1} + (J+\kappa) m_{n+1}^2 - \right.$$

$$\left. - 2H\theta_n m_n + J(\theta_n - \theta_{n+1})^2 + 2(k + 3\kappa)\theta_n^2 + \kappa\theta_{n+1}^2 \right], \qquad (5)$$

with the value $n = 0$ indicating the end dimer, and in the presence of extra spins, $\Delta W$ describes their interaction with the end dimer. The connections between variables $\theta_n$ and $m_n$ for any value of $n$ can be found from the equations $\partial W / \partial \theta_n = 0$, $\partial W / \partial m_n = 0$. For $n > 0$, there is an infinite set of equations having the same structure as those for the infinite ladder

$$J(2\theta_n - \theta_{n+1} - \theta_{n-1}) + (2k + 3\kappa)\theta_n - Hm_n = 0 \ , \qquad (6)$$

$$(4J + 2k + 3\kappa) m_n + J(m_{n+1} + m_{n-1}) - H\theta_n = 0 \ .$$

These equations can be solved using the exponential ansatz,

$$\theta_n = \theta_p e^{-np} + (-1)^n \theta_q e^{-nq} , \qquad (7a)$$

$$m_n = m_p e^{-np} + (-1)^n m_q e^{-nq} \qquad (7b)$$

where $p$ and $q$ are determined by substitution of (7) into Eqs. (6) to give $2\cosh p = 3\tilde{J}/J - 1$ and $2\cosh q = 3\tilde{J}/J + 1$, where $3\tilde{J} = \sqrt{9J^2 + H_1^2 - H^2}$ . The connetions between the amplitudes $\theta_{p,q}$ and $m_{p,q}$ can be written as $\theta_p = A\, m_p$ , $\theta_q = m_q / A$ , where $A \approx H/6J \ll 1$ . Thus, for a slowly decaying exponent with $p \ll 1$ one obtains $m_p \ll \theta_p$ , and for the staggered exponent with fast decay the situation is the opposite. Next minimization of $\Delta W$ with respect to the parameters describing the non-regular part of ladder with given values of $\theta_0$ and $m_0$ for the dimer with $n = 0$ gives this energy as a quadratic form with variables $\theta_0$ and $m_0$ . Then the analysis of stability can be done using the equation for $n = 0$, which by use of the relations $\theta_0 = \theta_p + \theta_q$ , $m_0 = m_p + m_q = A\,\theta_p + \theta_q / A$ can be written as

$$(Je^p - J - \kappa)\theta_p - (Je^q + J + \kappa)\theta_q + \partial\Delta W / 2\partial\theta_0 = 0 \ , \qquad (8)$$

$$(Je^p + J + \kappa) A\theta_p - (Je^q - J - \kappa)\theta_q / A - \partial\Delta W / 2\partial m_0 = 0 \, .$$

These equations have non-trivial solutions only for some particular value of the magnetic field, which defines the instability field, $H_c$ . Finally the critical field, as well as the connections between variables, $\theta_p$ and $\theta_q$. can be obtained through the solvability condition for Eqs. (8).

Using this procedure for the regular ladder ($\Delta W$ = 0), the critical field to a first approximation in the small parameters is

$$H_c^2 = H_1^2 - 2\left(K + \kappa\sqrt{3}\right)^2 . \tag{9}$$

Notice that this value is smaller than the volume critical field, where the difference between $H_c$ and $H_1$ is proportional to the small parameter $(\kappa, k)/J$ and for given effective anisotropy $K$ it is an increasing function of κ. This is in good agreement with numerical data illustrated in Fig. 2. It is also noticed that $\theta_q << \theta_p$, but $m_p$ and $m_q$ are comparable, $m_q \approx 2/(1+\sqrt{3})m_p \approx 0.73m_p$, implying that the oscillatory part of $m$ is quickly decaying, but observable as seen in Fig. 1.

Next this method is used to analyze the NC ladder. The minimization of $\Delta W$ with respect to the extra spin will give the simple expression for this energy in terms of $\theta_0$ and $m_0$ ,

$$\Delta W = -\alpha(\theta_0 - m_0)^2/2, \quad \alpha = H - 2k - 2\kappa + H^2/J . \tag{10}$$

Then the solvability condition for (8) gives the critical field $H_c^2 = 2H_1^2/5$, which is significantly below $H_1$ . For this case again $\theta_q << \theta_p$ and $m_p \approx m_q$ , but with opposite signs, $m_q \approx -m_p/(1+\sqrt{3}) \approx -0.37m_p$, which is also seen in the insert to Fig. 1.

Previously only first order spin flop transitions are known to exist for the model Hamiltonian given by Eq. (1) and only first order transitions were previously reported [6-8] in the spin chain literature. The order of the transition can be clearly demonstrated through a plot of $S_z$ versus $H$ , which goes to zero near the critical field for the second order transition. As was mentioned above, the first order transition appears for the RL and the NC ladders with pure single-ion anisotropy (see Fig.2), however, for the case of exchange anisotropy the transition becomes second order and the amplitude of the nonuniform distribution goes to zero at the transition point. Therefore, for some finite value of the ratio of the exchange and single-ion anisotropy constants there is a tricritical point. Taking into account that the magnetization per dimer, $m_{z,n} = \theta_n m_n$ is quadratic over linear variables, and that the value of $S_z$ = 0 in the Neel state, it is remarked that $S_z$ is the square of the order parameter. For description of the phase diagram with a tricritical point one needs a Landau expansion of the system energy containing the order parameter to the sixth power

$$F = -(H - H_c)S_z + \frac{1}{2}\beta S_z^2 + \frac{1}{3}\gamma S_z^3, \tag{11}$$

where the first term has been obtained analytically from the linear approximation, with β and γ being phenomenological coefficients. If the coefficients are known, then further analysis is simple: for $\beta > 0$ the transition is second order with an $S_z$ magnetic field dependence $S_z = (H - H_c)/\beta$. For $\beta < 0$ and $\gamma > 0$ the transition will be first order with nonzero value of total spin at the transition point given by $S_z^t = 3|\beta|/4\gamma$. Therefore, both quantities $(dS_z / dH)^{-1} = \beta$ for β > 0, and $S_z^t$ at β < 0 are proportional to the value of $|\beta|$. Near the tricritical point $\beta \to 0$ and plots of these quantities versus *k*/*K* at a constant value of the effective uniaxial anisotropy *K* = 2*k*+3κ can be extrapolated to zero to obtain the anisotropy value at the tricritical point. This graphical analysis for the NC ladder is illustrated in Fig. 3 and gives a tricritical point at the anisotropy ratio 2*k* / *K* ≈ 0.966. For the RL structure the tricritical point is determined to be at 2*k*/*K* ≈ 0.78, and the characteristic square root dependence is also seen in Fig. 2. Thus, the second order transition takes place for almost all values of anisotropies with *k* > 0 and κ > 0. For a small AFM particle bounded on all sides the situation becomes more complicated. It is more or less obvious the edge noncompensated spins are the main sources of pinning for nonuniform states. Use of numerical methods only for square particle with NC edge spins, it is remarked that the transition is always second order for both single-ion and exchange anisotropy.

These analytical and numerical calculations show that there is a rich array of ground state structures in AFM systems of dimensionality greater than one. In addition to the bulk spin flop state there are localized surface states of both the spin flop and discommensuration type depending on whether or not the surface is compensated. The transition to these states can be either first or second order, which can be determined from the $S_z(H)$ behaviour near the transition point. The second order transition is realised for all systems considered with exchange anisotropy, as well as for the 2D model of a square particle with NC edge spin. As the second order transition is approached, the amplitude of the non-uniform spin distribution goes to zero at the transition point, producing a state, which is neither SSF nor discommensuration, rather it is a slightly broken Neel state. For spin ladder models, RL and NC structures with combined anisotropy, a tricritical point on the (*H*, *k*/*K*) plane is present.

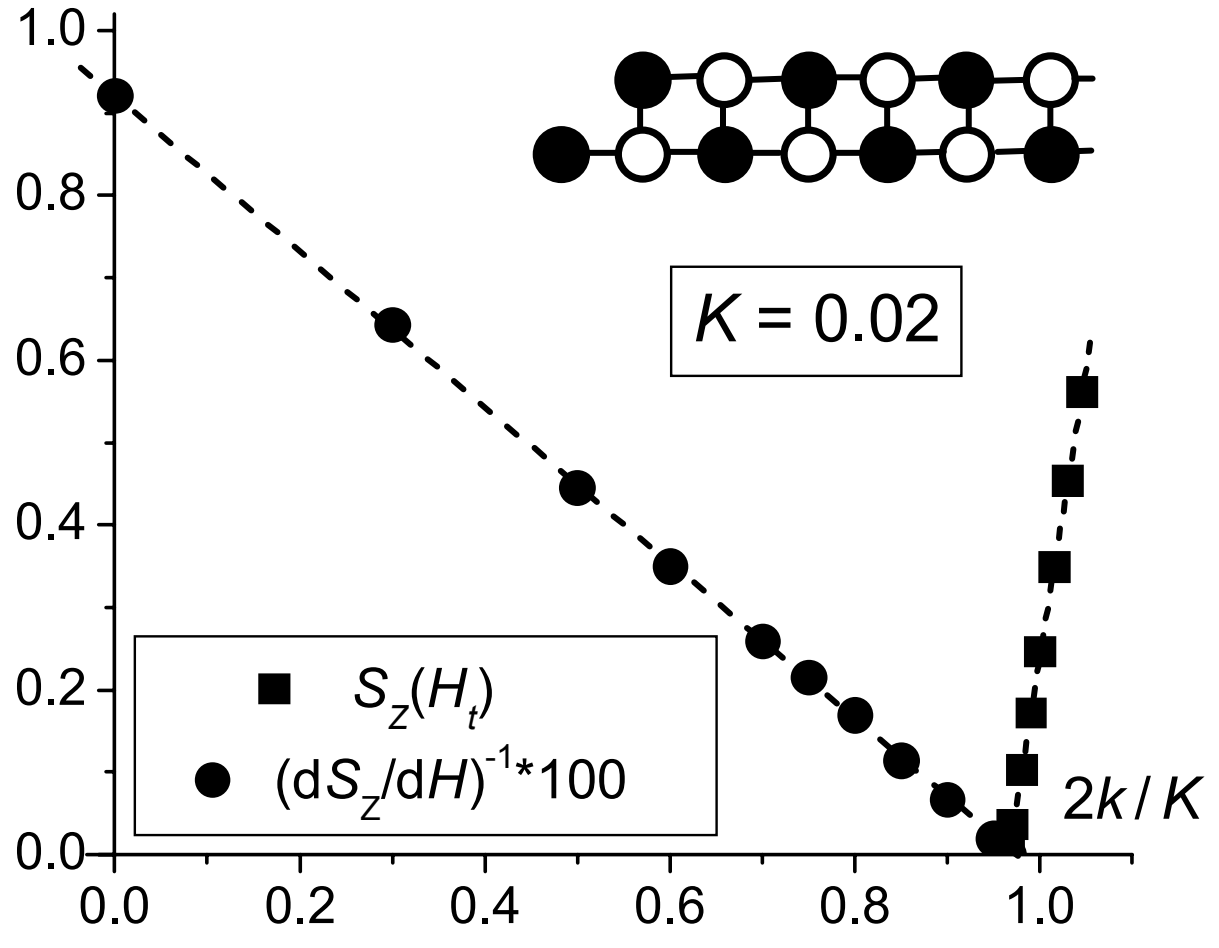

Fig.3. Plots of $S_z$ and $\left(dS_z/dH\right)^{-1}$ versus $k/K$ for $K=0.02$ on a NC ladder shown in the inset.

The two-dimensional square planar structure can be realized in layered classical AFM like Mn(II)-halide compounds which can be approximately modeled by a classical spin [9]. Another interesting possibility is provided by arrays of nanostructures such as magnetic dots [10], which can be synthesized in square planar or ladder structures. Owing to the magnetostatic interaction between the dots, these have AFM ordering of the dot's magnetic moment at zero field [11] . These effects could possibly be observed in these magnetic dot arrays, where the effective anisotropy for one dot as well as the anisotropy of dots interaction (the analogue of single-ion or exchange interaction) can be adjusted by modification of the geometries of the dot and lattice, respectively.

ACKNOWLEDGMENTS

The authors thank A.K. Kolezhuk for fruitful discussions. This research was supported by the National Science Foundation grant number DMR-9974273, and the work in Kiev was supported financially by a Volkswagen Stiftung grant no. I/75895.